\documentclass{emulateapj}

\shorttitle{Anelastic and Compressible Oxygen Burning}
\shortauthors{Meakin \& Arnett}

\def \nuc#1#2{\relax\ifmmode{}^{#1}{\protect\text{#2}}\else${}^{#1}$#2\fi}

\def\mcol{\multicolumn}
\def\mult{$\times$}
\def\msun{M$_{\odot}$ }
\def\lang{\langle}
\def\rang{\rangle}

\begin{document}
  \title{Anelastic and Compressible Simulations of Stellar Oxygen Burning}    
  \author{Casey A. Meakin\altaffilmark{1} \& David Arnett\altaffilmark{1}}
    \altaffiltext{1}{Steward Observatory, University of Arizona,  
      933 N. Cherry Avenue, Tucson, AZ 85721}
    \email{cmeakin@as.arizona.edu,darnett@as.arizona.edu}

  \begin{abstract}
    In this paper we compare fully compressible \citep{ma06a,ma06b} and anelastic \citep{kwg03} 
    simulations of stellar oxygen shell burning.  It is  found that the two models are in agreement 
    in terms of the velocity scale ($v_c \sim 10^7$ cm/s) and thermodynamic fluctuation amplitudes 
    (e.g., $\rho'/\lang\rho\rang \sim 2\times10^{-3}$) {\em in the convective flow}. Large fluctuations 
    ($\sim$11\%) arise in the compressible model, localized to the convective boundaries, and are 
    due to internal waves excited in stable layers.  Fluctuations on the several percent level are 
    also present in the compressible model due to composition inhomogeneities from ongoing entrainment 
    events at the convective boundaries.  Comparable fluctuations (with amplitudes greater than $\sim$1\%) 
    are absent in the anelastic simulation because they are due to physics not included in that model. 
    We derive an analytic estimate for the expected density fluctuation amplitudes at convective boundaries 
    by assuming that the pressure fluctuations due to internal waves at the boundary, $p_w'$, balance 
    the ram  pressure of the convective motions, $\rho v_c^2$.  The predicted amplitudes agree well with 
    the simulation data.  The good agreement between the anelastic and the compressible solution within 
    the convection zone and the agreement between the stable layer dynamics and analytic solutions to the 
    non-radial wave equation indicate that the compressible hydrodynamic techniques used are robust for 
    the simulated stellar convection model, even at the low Mach numbers found $M\sim0.01$.
  \end{abstract}


\section{INTRODUCTION}
Oxygen burning (by $\rm ^{16}O+ ^{16}O$ fusion) occurs in the precollapse
stages of the evolution of massive stars. Neutrino cooling speeds 
these stages to the extent that the evolutionary times scales are
close enough to the sound travel time so that direct compressible
numerical hydrodynamics can be applied \citep{da94}. The first
detailed studies of this stage \citep{ba98} were done in two-dimensional
symmetry (2D) with PROMETHEUS \citep{fma89}, a multi-fluid multidimensional
compressible hydrodynamics code based on the Piecewise Parabolic Method
(PPM) of \citet{cw84}. They showed vigorous convection, with significant
density fluctuations (up to 8\%) at the convective-nonconvective boundaries. 
These results were confirmed in detail in 2D with the VULCAN 
Arbitrary-Lagrangian-Eulerian (ALE) 
hydrodynamics code \citep{eli93} by \citet{sa00}.
VULCAN is an entirely independent compressible hydrodynamics code,
so that these two sets of simulations only shared the initial model, the sonic
time step limitation, and the 2D geometry. 
A new version of the PROMETHEUS code, PROMPI (which uses the Message
Passing Interface for parallelism), has extended the study
to 3D. In all these compressible models except the earliest \citep{da94} the 
computational domain has included both the convective oxygen burning shell as 
well as two bounding stably stratified layers.

\citet{kwg03} investigated shell oxygen burning in 3D using an anelastic
hydrodynamics code which filters out sound waves and linearizes thermodynamic
fluctuations around a background reference state
\citep{gg84,gough69}. 
In contrast to the fully compressible results above, \citet{kwg03} found only small 
density and pressure contrasts, and subsonic flows which were well within the anelastic 
approximation (all thermodynamic contrasts less than 1 percent).
The boundary conditions used were impermeable and stress-free and were placed
within the convection zone so that convective overshoot could not be studied. 
In particular, the dynamic consequences of the neighboring non-convective shells, 
and their elastic response to convective fluctuations, were ignored.
The formulation was single fluid, so that effects depending upon
composition, i.e., mixtures of fuel and ashes, were not modeled.

\par The applicability of both fully compressible and anelastic hydrodynamic
methods have recently been challenged by developers of low-Mach number solvers \citep{almgren06}.
The reliability of compressible codes has been questioned for low velocity flows
due to possible violations of elliptic constraints that arise in the evolution equations
in the very small Mach number limit \citep[e.g.,][]{schneider99}.  
The limits for which compressible solvers remain robust in the astrophysical context, 
however, has not been rigorously studied.  Anelastic methods, on the other hand, 
enforce a divergence constraint on the velocity field which filters out sound waves,
but are formulated assuming that thermodynamic fluctuations are small and only linear deviations
from a background reference state are retained.  Therefore, this approximation is expected to 
fail for models which include large gradients in the thermodynamic variables such as occur at 
the boundaries of shell burning regions.

The correct identification of the behavior in shell oxygen burning has wide 
implications.  This stage of massive star evolution is important for a variety of 
topics of current research interest (e.g., \citet{yetal05,yetal06}). 
In this paper we discuss the quantitative similarities and differences between
the models of oxygen shell burning which we have introduced above and show that
the two models are in good agreement with each other. In addition, the compressible
models are in agreement with scaling relations derived from the basic hydrodynamic 
equations as well as analytic solutions to the non-radial wave equation for motions
in the stable layers.
These findings lend strong support to the validity of both simulations.  
There are no signs that either the anelastic or the PPM method is breaking down for the 
conditions simulated, even in regions of the flow where the Mach number does not 
exceed $M\sim0.01$.

\section{MODEL COMPARISON}

\subsection{The Initial Models and Simulation Parameters}

\par In Table 1 we summarize the initial conditions, computational
domains, zoning, and properties of the developed flow for the three models that we 
will discuss in this paper.
These include a 2D and a 3D compressible model calculated with the PPM method \citep{ma06a,ma06b} 
and the non-rotating anelastic model described by \citet{kwg03}. The initial conditions
for the compressible simulations are of a 23 \msun star previously evolved with 
the TYCHO stellar evolution code \citep{ya05} which is directly mapped onto the hydrodynamics 
grid.  A 25 nuclei reaction network is used to track composition and energy generation.  
The computational domain for the compressible models are restricted to fractions of a sphere
and use a spherical coordinate system.  The 2D model is a 90$^o$ wedge embedded in the
equatorial plane, and the 3D model is a wedge of 30$^o$\mult30$^o$ degrees centered on the
equator.  The radial limits for these models enclose both the convectively unstable oxygen burning
shell, as well as two surrounding stably stratified layers.  Boundary conditions, which are
{\em placed in the stable layers}, are impermeable and stress free. The net energy generation
due to nuclear burning and neutrino cooling, 
L$_{net} = \int (\epsilon_{nuc}+\epsilon_{\nu\bar{\nu}}) dM \approx 3.5\times10^{46}$ erg/s, 
is positive and goes into PdV work through a background expansion which develops 
naturally in the compressible model in the same way it does in the initial TYCHO model.

\par The anelastic model uses a reference state which is a polytropic fit to a 25 \msun stellar 
model which was evolved with the KEPLER code \citep{wzw78}. 
A multi-fluid model and reaction network are not used and nuclear energy generation is instead 
estimated using a power law fit for density and temperature dependence. The anelastic model uses 
spherical harmonics for angular coverage, ameliorating the pole singularity
problem of a spherical coordinate system, and covers a full 4$\pi$ steradians.
Chebyshev polynomials are used in the radial direction. The radial limits of the anelastic model 
enclose {\em only the convection zone with no regions of stable  stratification}.  
The boundary conditions, which are within the unstable convective layer,
are also impermeable and stress free.  The net energy generation is positive, 
L$_{net}\approx$ 1.5\mult10$^{45}$ erg/s.  The anelastic model used is unable to model background
expansion so the excess energy is forced to escape from the outer
boundary of the calculation.
The lower net energy generation of this model may be due to it being in a
later evolutionary stage.

\par Both hydrodynamic models use the equation of state provided by \citet{ts00}.
The radial limits of the oxygen burning convection zone and enclosed 
mass for the two initial models evolved with the KEPLER and TYCHO codes are remarkably similar.  
The compressible model uses 400 logarithmically spaced radial zones 
(to keep zone aspect ratio $dr/rd\theta\sim$1) and an angular resolution of $\sim$0.3$^o$ per zone.  
The anelastic model uses 145 zones for a comparable radial extent, and spherical
harmonics up to order $l=63$ to cover the sphere which is roughly equivalent to a Nyquist sampling 
of $\sim$1.5$^o$ per zone, approximately a factor of five lower angular resolution than
the compressible model.  The Rayleigh and Reynolds numbers quoted by \cite{kwg03} are 
$Ra\sim5\times10^7$ and $Re\sim3000$.  Since the compressible model is more strongly 
driven (larger L$_{net}$) and has finer zoning (resulting in a lower effective viscosity), 
both the effective Rayleigh and Reynolds numbers will be higher in the compressible model.

\subsection{Flow Properties: Anelastic Model}

\par The anelastic simulation has been run for 6500 seconds.
With an average flow velocity
of $v_c \approx$ 0.49\mult 10$^7$ cm/s, and a radial extent for the convection 
zone $\Delta R \approx$ 0.39\mult10$^{9}$ cm the turnover time 
$t_c=2 \Delta R/v_c \approx$ 159 seconds and the simulation spans approximately 41
convective turnovers.  The peak velocity is given as $v_{peak} \approx 1.8$\mult 10$^7$ cm/s
which corresponds to a peak Mach number $M\sim0.04$ for a sound speed $c_s \approx 4.5$\mult 10$^{8}$ cm/s.
The maximum density fluctuations within the convection zone are found to be of the order
$\rho'/\langle\rho\rangle \sim 2\times10^{-3}$ which is the same order of magnitude as 
the peak Mach number squared, $M^2 \sim 1.6\times10^{-3}$ consistent with the 
scaling arguments for the anelastic approximation for thermal convection \citep{gough69}.

\subsection{Flow Properties: Compressible Models}

\begin{figure*}[t]
  \plotone{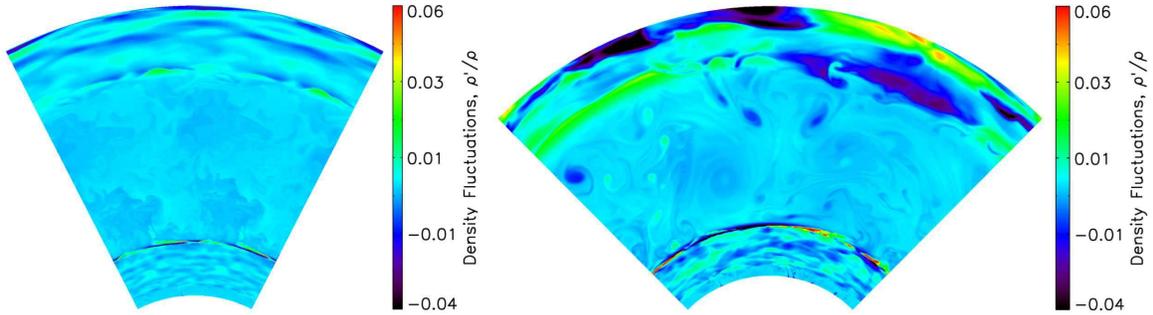}
  \caption{\footnotesize
    Spatial distribution of density fluctuations are shown for 
    (left) a slice through the 3D compressible model and (right) the 2D 
    compressible model.  In both models the flow is composed of two distinct regimes,
    including the convective flow in the center which is bounded above and below by
    stably stratified layers which are host to internal waves.
    The scale on the 2D model has been truncated to same limits as the 3D model
    for comparison, but extreme values exceed the scale limits by a factor of $\sim$2.
    The 3D model has been tiled twice in angle for clarity.
    \label{fig_1}} 
\end{figure*}

\par The flow in the compressible simulations consists of two distinct regimes: the
turbulent convection zone, and the wave-bearing stably-stratified layers. These can be quite readily
discerned in the density fluctuation field shown in Figure \ref{fig_1}.
In this section we discuss these two regimes in turn for the three dimensional model, 
and then discuss the properties of the two dimensional model.

\par
\begin{figure*}[t]
\plottwo{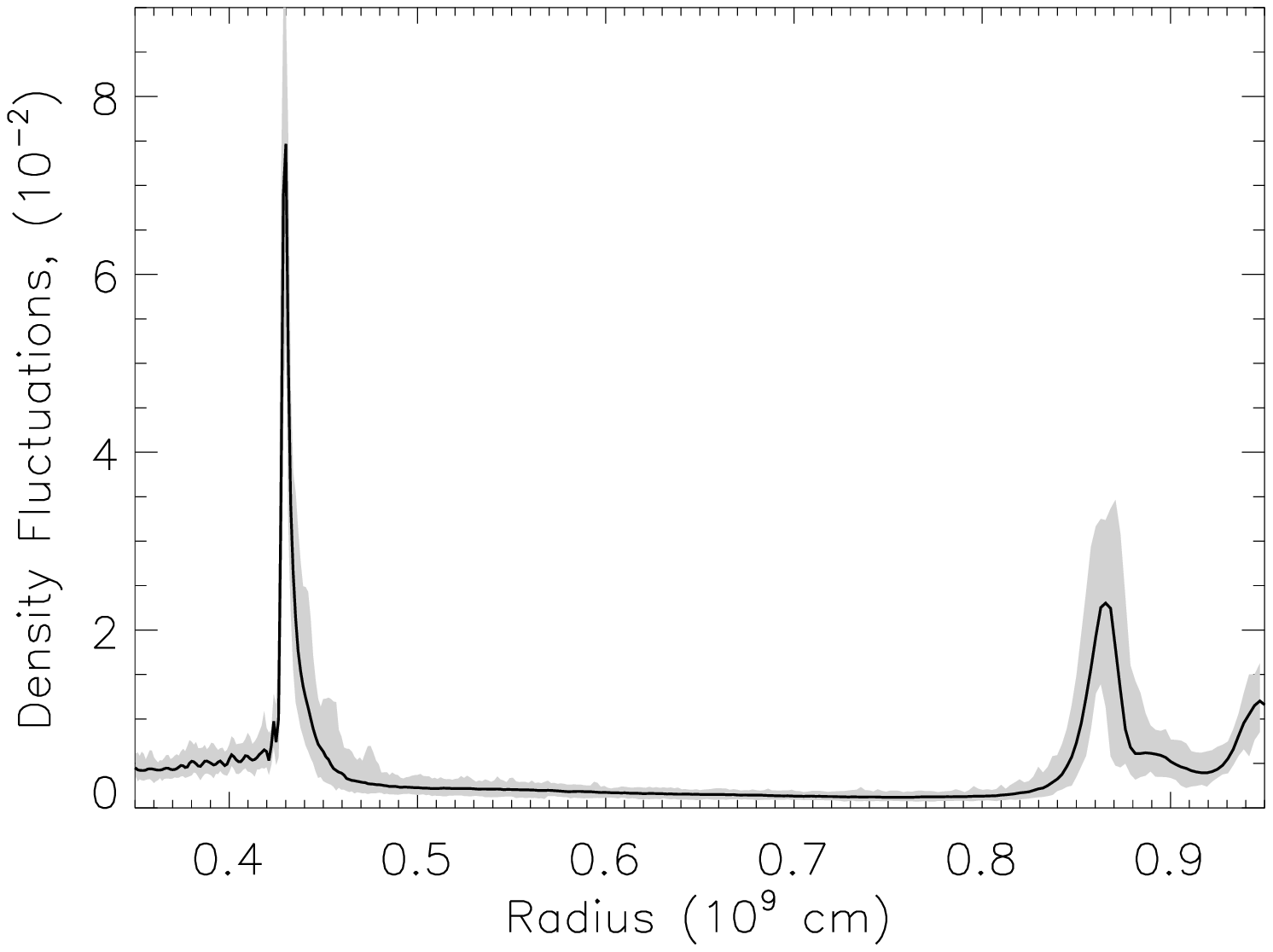}{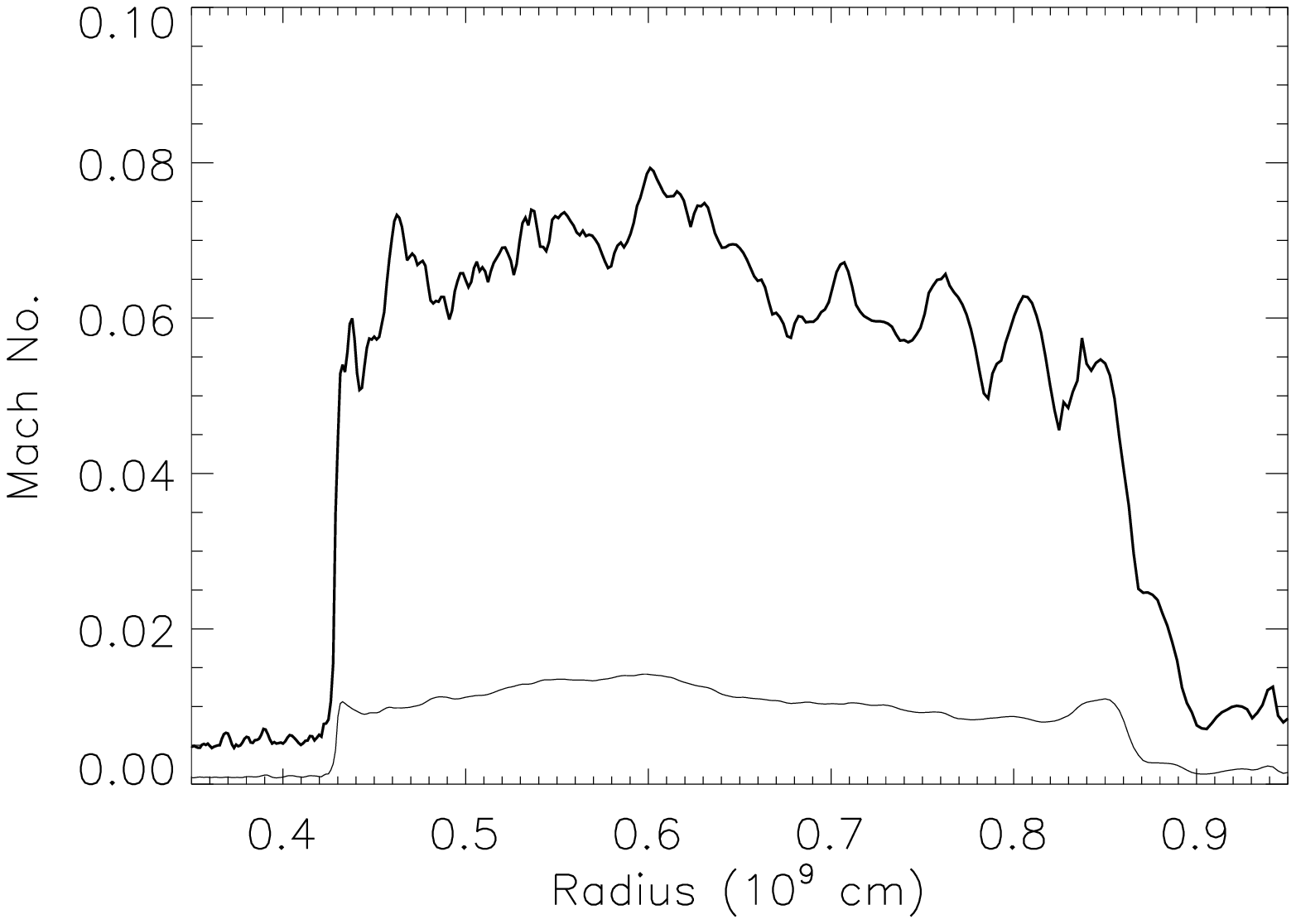}
\caption{\footnotesize
  (left) Density fluctuations for 3D compressible model: The time average of the
  maximum density fluctuation is shown as the thick line, with the extreme values
  for the averaging period (two convective turnovers) shown as the shaded region.
  Within the body of the convection zone the average fluctuations are quite low $\sim$0.25\%.
  The largest fluctuation (the spike at $r \approx 0.43 \times 10^9 \rm\ cm$) 
  is $\sim$11\% (exceding the plot limits),  while the secondary 
  maximum ($r \approx 0.85 \times 10^9 \rm\ cm$) reaches $\sim$3.5\%. 
  These fluctuations occur in the radiative regions that enclose the convection  zone 
  (outside the computational domain of \citet{kwg03}).
  (right) Mach number for the 3D compressible model: shown is the instantaneous value of
  the maximum Mach number at a given radius (thick line), and the rms Mach 
  number (thin line).
  \label{fig_2}}
\end{figure*}

 The three dimensional compressible model was run for $\sim$800 seconds of star time.  
The average flow velocity is found to be $v_c \approx$ 0.8\mult10$^7$ cm/s.  With a convection 
zone width $\Delta R \approx$ 0.41\mult10$^9$ cm the turnover time is $t_c \approx $ 103 seconds
and the simulation spans approximately 8 convective turnovers. 
After an adjustment in the initial size of the convection zone due to penetrative 
convection \citep{ma06a,ma06b}, the flow achieves a steady state within $\sim$200 seconds, or two 
convective turnovers, after which the average flow properties do not change appreciably.
Figure \ref{fig_2} shows the peak density fluctuation and the Mach number at each radius
for the three dimensional model.
Within the convection zone, 0.44 $\la$ r/10$^9$ cm $\la$ 0.85, the 
maximum density fluctuation and Mach number are  
$\rho'/\langle\rho\rangle \sim 5\times10^{-3}$ and $M\sim$ 0.09, 
respectively.  Here also, the fluctuation scale is the same order of magnitude as the
peak Mach number squared, $M^2 \sim 8\times10^{-3}$.
The rms Mach number in the convection zone is $M_{rms}\sim$0.01.

\par These velocity and fluctuation scales are comparable to those of the 3D anelastic 
model and are listed in Table 1 for both simulations for comparison.
The point we want to emphasize here is that {\em the character of the convective flow is 
  quantitatively in agreement between the anelastic and compressible models.} 
We also find, from a detailed comparison to be presented separately \citep{ma06b},
that the 3D compressible model compares well with the stellar mixing length 
theory \citep{kw90}, including the velocity scale ($v_c\sim 10^7$ cm/s) and the
superadiabatic stratification ($\Delta\nabla\sim 5\times10^{-4}$).  The slightly
larger velocity scale in the compressible model can be attributed to the higher Rayleigh 
number due to the larger luminosity needed to be transported by the convective flow.

\subsubsection{Additional Sources of Fluctuations}

\par In this section we discuss the origins of thermodynamic fluctuations
which are due to physics not included in the anelastic model
of \cite{kwg03}, including stably stratified layers and composition
effects.  We begin with a discussion of the internal wave dynamics
occurring near the convective boundaries.

\par Significant density fluctuations occur at the convective boundaries 
in the compressible model, reaching values as large as 11\mult10$^{-2}$,
over twenty times larger than in the body of the convection zone
(Figure \ref{fig_2} and Table \ref{tab_1}). 
Examining the spatial distribution of the density fluctuations presented in Figure \ref{fig_1}
reveals that the largest fluctuations occur at the interface between the convection zone
and the stably-stratified, wave-bearing layers.  The morphology of the largest density
fluctuations in the domain (i.e., those at the convective boundary) are periodic in angle 
and harmonic in time, identifying them with internal wave dynamics.
In the following discussion we present analytic estimates for the amplitudes of the
density fluctuations at the convective boundary and show that they are in good agreement 
with the simulation data.  

\par For small amplitude waves the Eulerian density fluctuations and pressure fluctuations 
are related by \citep[p.93]{unno89}:

\begin{equation}
  \frac{\rho'}{\lang\rho\rang} = \frac{p'}{\lang p\rang}\frac{1}{\gamma_{ad}} +
  \xi_r \frac{N^2}{g} 
  + \hbox{(nonadiabatic terms)}
  \label{eq_rhop}
\end{equation}

\noindent with buoyancy frequency $N$, and Lagrangian displacement $\xi_r$.
We defer a discussion of nonadiabatic and composition effects to the end of this section.
In the convection zone, material is nearly neutrally stratified and the
buoyancy frequency is very close to zero so the second term on the right hand side 
is not very important.  At convective boundaries the stellar
structure assumes a stable stratification with a positive buoyancy
frequency, and this term can become dominant. 
This term represents the component of the Eulerian density fluctuation
due to g-mode oscillations and is the projection of the Lagrangian displacements of the wave, $\xi_r$, 
onto spherical shells. In the presence of steep density gradients, waves can lead to large 
Eulerian fluctuations even when compressibility is not important. To be as clear as possible 
on this key point, we give an analogy to well known physics: consider waves on a lake. The 
Lagrangian surface is the surface of the water. The Eulerian surface is the average level
of the water and Eulerian density fluctuations occur as the waves (water) and
troughs (air) move by the observer. The large variation in density is
not due to compression, but the choice of coordinates (Eulerian
in this case).

\par In order to estimate the amplitude of the density fluctuations using equation \ref{eq_rhop}
we need to know the size of the pressure fluctuation and the maximum radial displacement amplitude, 
$\xi_r^{\hbox{\tiny max}}$, for wave motions at the boundary.  Both quantities can be estimated by 
assuming that the ram pressure of the convective turbulence is balanced by wave induced pressure 
fluctuations at the convective/stable layer interface:

\begin{equation}
  \rho v_c^2 \approx p_w'.
  \label{pbalance}
\end{equation}

\begin{figure*}[t]
  \plottwo{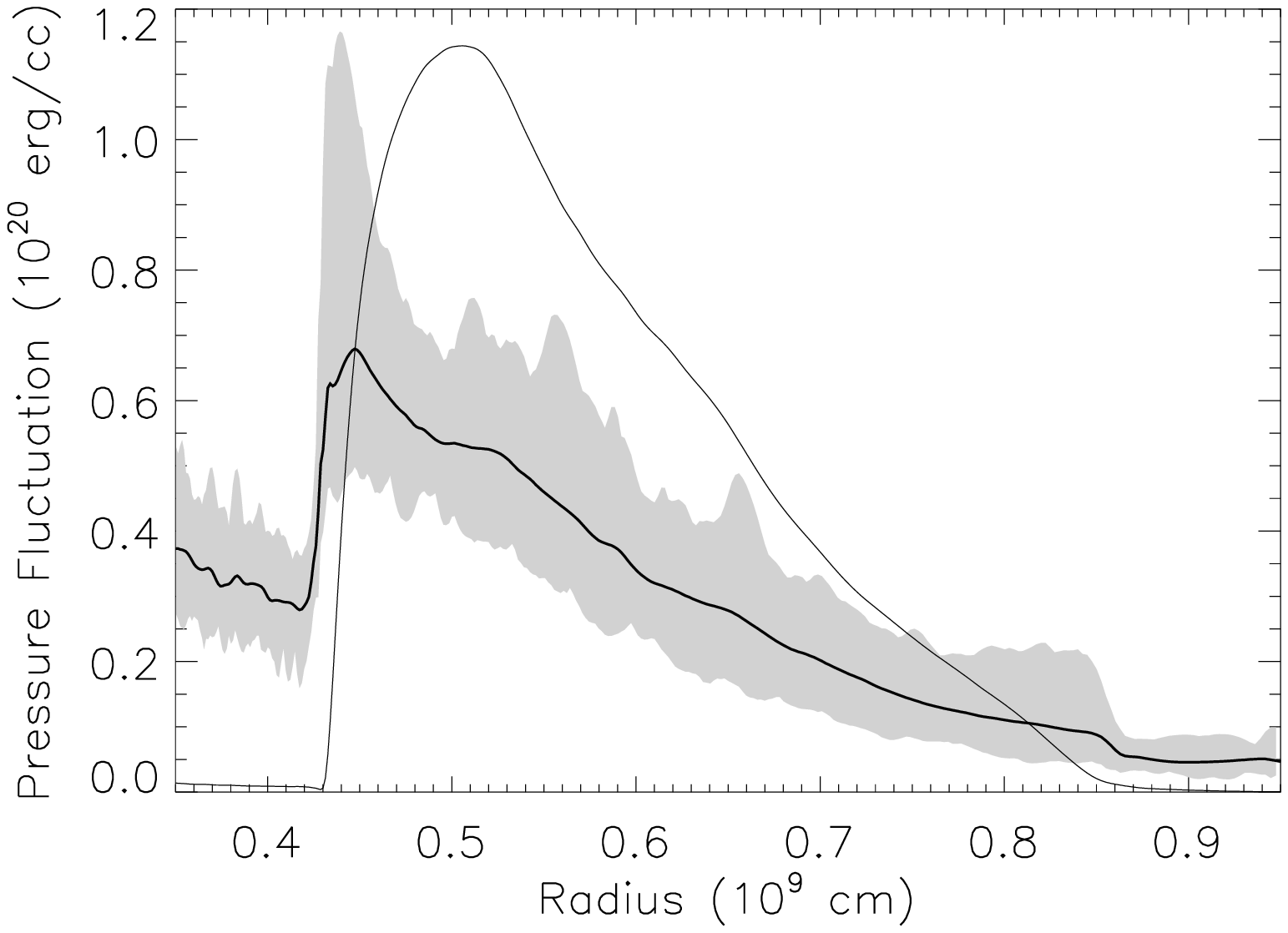}{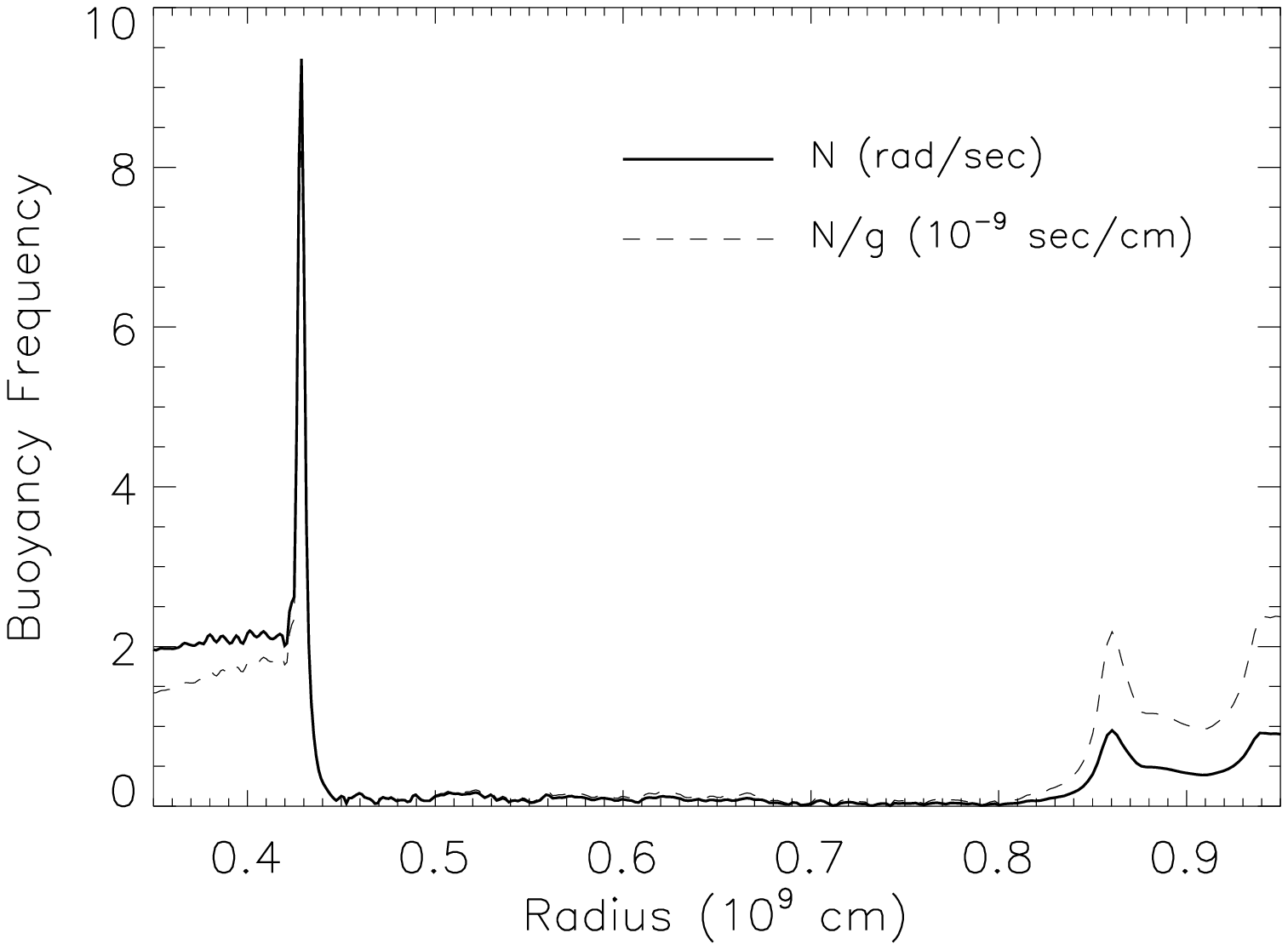}
  \caption{\footnotesize
    (left) Pressure fluctuations in 3D compressible model: 
    The time averaged horizontal RMS pressure fluctuations are shown 
    as the thick line, with the envelope of extreme values over two 
    convective turnovers indicated by the shaded region.
    The radial ram pressure of the turbulent convection, $\rho v_r^2$,
    is shown as the thin line.  The curves cross at the convective boundaries where
    the turbulent pressure is balance by the pressure fluctuation induced by internal
    waves in the adjoining stably stratified layers.
    (right) The buoyancy frequency is shown in units of rad/s.  Also shown as the
    dashed line is the buoyancy frequency normalized by the gravity which,
    through Equation~\ref{dpert}, sets the scale of the density fluctuations at the
    convective boundaries (compare with Figure \ref{fig_2}a).
    \label{fig_3}}
\end{figure*}

\noindent The validity of this approximation is demonstrated in Figure \ref{fig_3},
which shows that the RMS horizontal pressure fluctuations and the turbulent ram pressure
are comparable in the convection zone and do indeed balance at the locations of the 
convective boundaries.  

\par The relationship between the pressure fluctuation of an internal 
wave, $p_w'$, and the maximum radial displacement, $\xi_r^{\hbox{\tiny max}}$, depends on 
the wave frequency $\sigma$ and horizontal wavenumber $k_h$.
Perhaps the simplest approximation is to assume that internal waves
generated at the convective
boundary are directly related to the convection through the convective velocity $v_c$ and 
eddy scale, $l_c \sim H_p$, by $v_c \sim \sigma/k_h$ and 
$k_h \sim 2\pi/l_c$ in the spirit of stellar mixing length theory \citep{press81}. 
Adopting these values we can then use the linearized momentum equation
\citep[p.96]{unno89},

\begin{equation}
  \xi_h^{\hbox{\tiny max}} \sim \frac{[l(l+1)]^{1/2}}{r\sigma^2}\frac{p_w'}{\rho} 
  \approx \frac{k_h}{\sigma^2} v_c^2
\end{equation}

\noindent and the dispersion relation (for waves in which $\sigma \ll N < L_l$), 

\begin{equation}
  k_h/k_r \sim \frac{\sigma}{N}
  \label{dispersion}
\end{equation}

\noindent to estimate the radial displacement:

\begin{equation}
  \xi_r^{\hbox{\tiny max}} = \xi_h^{\hbox{\tiny max}}\times k_h/k_r 
  \sim \frac{k_h v_c^2}{\sigma^2}\frac{\sigma}{N} \sim v_c/N.
  \label{rdist}
\end{equation}

\noindent The dispersion relation used to connect the horizontal and
radial displacement amplitudes is valid when the wave frequency
is much smaller than both the buoyancy frequency and the Lamb frequency
$L_l = k_h c_s$, with sound speed $c_s$, and is a reasonable
approximation for the wave properties adopted above 
(where $\sigma/L_l \approx M_c$).

\par Here, our main result is the last expression in equation \ref{rdist} for the
radial displacement amplitude of the interfacial wave which is in pressure balance
with the ram pressure of the convection. This expression is equivalent to the statement
that the kinetic energy of the turbulent motion exciting the wave is balanced by the
potential energy of the wave, which follows naturally from the basic energetic properties 
of waves in fluids \citep{lh78}. 

\par Finally, we use the displacement given by equation \ref{rdist}
and the pressure fluctuation in equation \ref{pbalance} with 
equation \ref{eq_rhop} to arrive at our estimate for the interfacial 
density fluctuation amplitude,

\begin{equation}
  \frac{\rho'}{\lang\rho\rang} \sim M_c^2 + \frac{v_c N}{g}
  \label{dpert}
\end{equation}

\noindent in terms of the Mach number $M_c$, gravity $g$, and buoyancy frequency N.
Adopting flow parameters from the simulation ($v_c ~\sim 10^7$, $g\sim 10^9$ in cgs units)
we find $\rho'/\lang\rho\rang \sim (10^{-3} + N \times 10^{-2})$ with $N$ in 
rad/s.  The validity of this expression is apparent when comparing the buoyancy frequency
in Figure \ref{fig_3} with the density fluctuations at the convective boundaries in Figure
\ref{fig_2}.

\par It can also be seen that the density fluctuations throughout the stable regions,
not just at the convective boundaries, are in rough agreement with the scaling given by
equation \ref{dpert}, though the amplitudes in the stable layers drop off with distance
from the convective boundary.
This is due to a spectrum of internal wave modes excited at 
the convective boundaries, rather than the single mode assumed in the above analysis.
Each modal component contributes to the pressure balance at the convective boundary
and is composed of wave packets that travel back and forth within the resonating
cavity of the stable layer \citep{unno89} causing fluctuations throughout the region.
A more detailed analysis is possible in which the entire spectrum of
internal waves is estimated by matching the wave motions to those of the spectrum of turbulent 
convection \citep[e.g.][]{carruthers86}.  Our single mode approach, however,
works very well in describing the amplitudes of the fluctuations at the convective 
boundaries and provides a reasonable upper limit to the fluctuations throughout the
entire stable layer.

\par We conclude this section with a discussion of the role that entropy fluctuations
play  in setting the scale of density fluctuations and hence buoyancy 
of material in the convection zone.  The non-adiabatic term in equation \ref{eq_rhop} takes 
the form, 

\begin{equation}
  \frac{\rho'}{\lang\rho\rang} = 
  \frac{v_T}{c_p}\delta S
\end{equation}

\noindent with thermodynamic derivative $v_T = -(\partial\ln\rho/\partial\ln T)_p$, 
specific heat at constant pressure $c_p$,
and Lagrangian entropy fluctuation $\delta S$.  In the present model the largest
non-adiabaticity is due to net effect of nuclear burning and neutrino cooling.
The entropy fluctuation can then be written 
$\delta S_{nuc} = \delta Q_{net}/T \approx \epsilon_{net}\delta t/T$ where
the time material dwells in the burning region is 
$\delta t \approx \Delta r/v_c \sim l_c/v_c \sim 10 s$.  
For the nuclear energy release in the current model, with a peak burning rate of 
$\epsilon_{net}\sim 10^{14}$ erg/g/s, %
the maximum entropy fluctuation will be of order $\delta S_{nuc}^{max} \sim 0.01$ in units
of $N_A k_B$.  The corresponding maximum density fluctuation
will be approximately $\rho'/\lang\rho\rang|_{nuc} \sim 2.5\times10^{-3}$, which is
of order the amplitude of the fluctuations due to the weak compressibility effects in 
the convective flow.  These values compare well to the perturbations estimated using mixing
length theory, consistent with the convective flow being driven by the nuclear burning
luminosity in the shell.

\begin{figure*}[t]
  \epsscale{0.95}
  \plotone{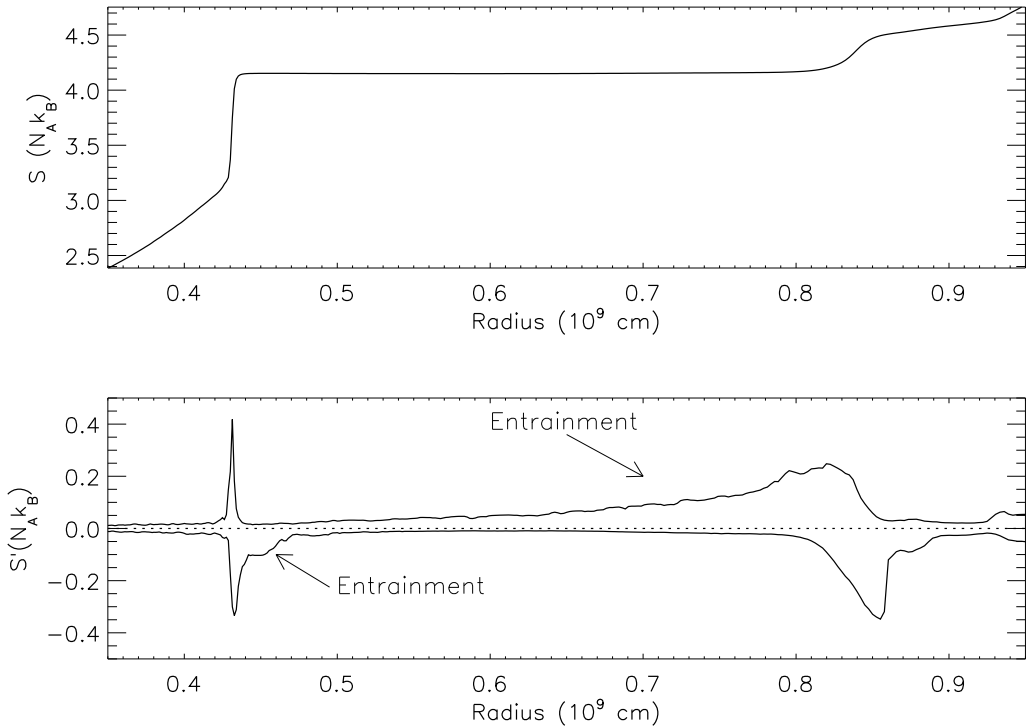}
  \caption{\footnotesize
    (top) Entropy profile of for 3D compressible model.
    (bottom) Entropy fluctuations for 3D compressible model:  The two solid lines indicate the
    maximum and minimum  fluctuation for a given radius over the course of two convective turnovers.
    The annotations indicate fluctuations due to low entropy material entrained at the
    lower boundary and high entropy material entrained at the upper boundary.
    \label{entropy}}
\end{figure*}

\begin{figure*}
  \epsscale{0.95}
  \plotone{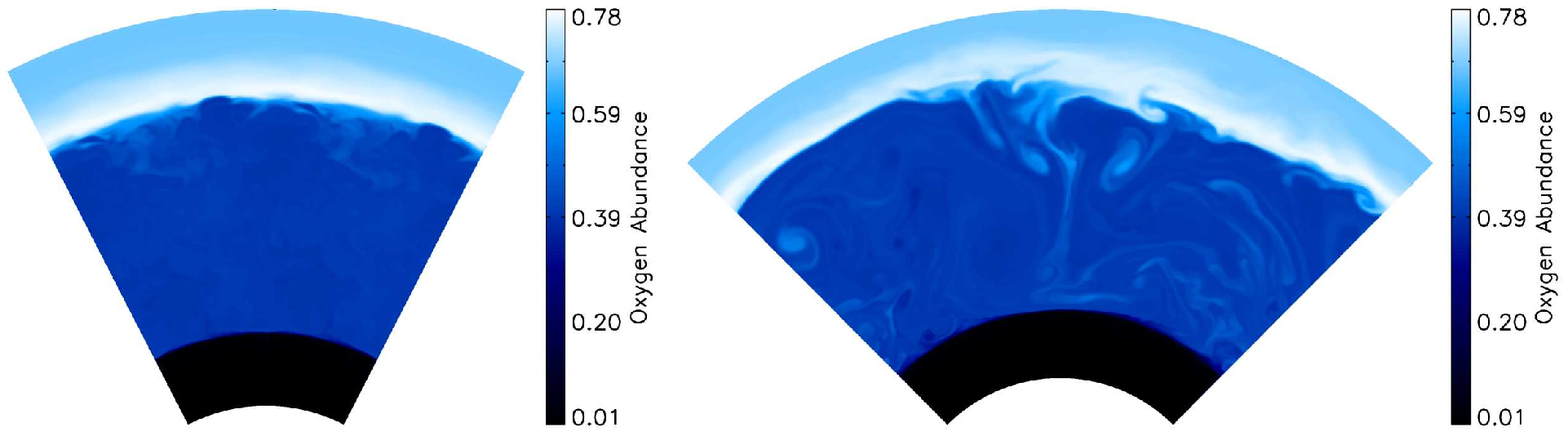}
  \plottwo{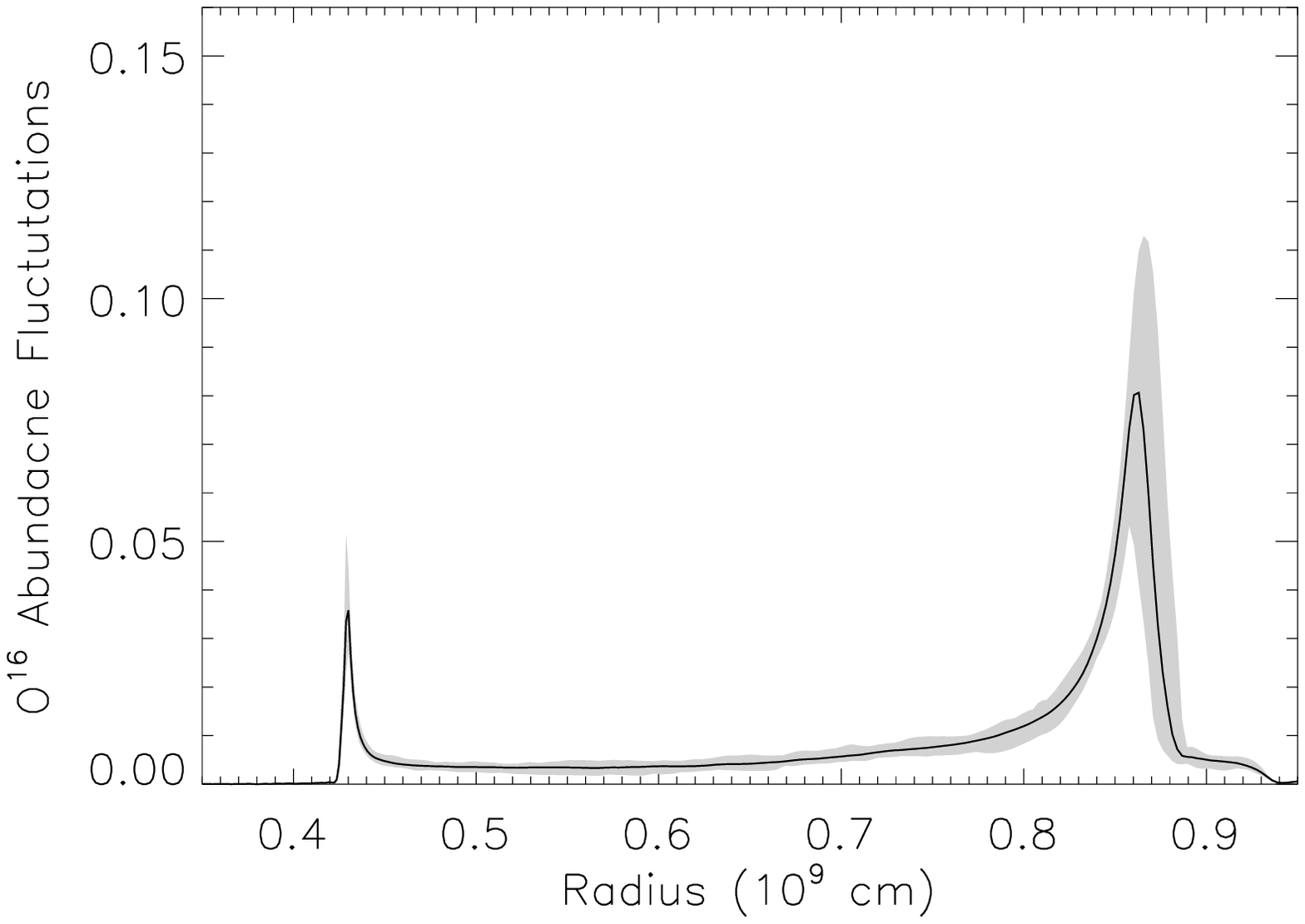}{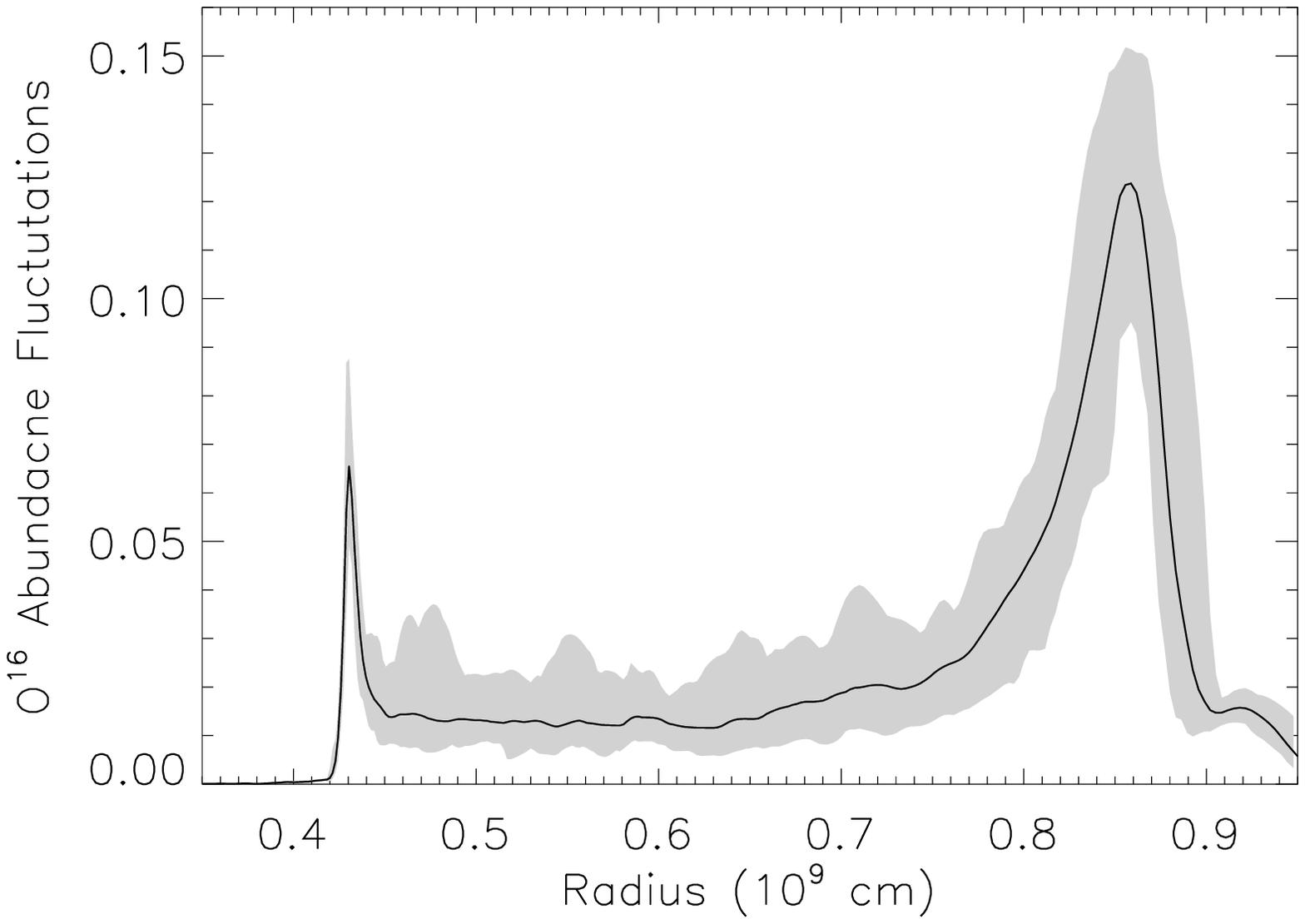}
  \caption{\footnotesize
    Spatial distribution of the oxygen mass fraction is shown 
    for (left panels) the 3D compressible model and (right panels) the 2D 
    compressible model.  The spatial distribution is shown in the top
    row, and the time averaged RMS horizontal fluctuations are shown in the bottom
    row wih the shaded region indicating extreme values of the fluctuations over two 
    convective turnovers.
    \label{composition}}
\end{figure*}

\par Entropy fluctuations also occur within the convection zone due to composition
inhomogeneities. %
The radial entropy profile and the RMS entropy 
fluctuations are presented in Figure \ref{entropy}.
The fluctuations are due to: (1) interfacial wave motions which cause 
Eulerian fluctuations in the same manner as for the density fluctuations discussed 
above; and (2) the entrainment of high and low entropy material at the convective boundaries 
which is mixed into the nearly adiabatic convection zone.  The wave induced fluctuations
appear as spikes near the convective boundaries and are present in both the
curve of minimum and maximum fluctuation.
The regions affected by compositional inhomogeneities are labeled 
in Figure \ref{entropy}, with low entropy 
material entrained from below and high entropy material entrained from above.
The entropy fluctuations associated with this material is another source of density 
fluctuations and explains the larger values that occur just within the boundaries
of the convection zone in Figure \ref{fig_2}.  The entropy fluctuations associated with
the entrained material are much larger than due to nuclear burning (the entropy 
perturbation in the region of greatest nuclear energy deposition,
r $\sim 0.45\times 10^9$ cm, is primarily {\em negative}).
The entrainment of material from stable layers by a turbulent convective flow
is an essential addition to stellar evolution modeling with significant consequences
for the evolution of burning shells in presupernova models.

\subsubsection{Dimensionality}

\par It has long been known that 2D simulations were informative only to the extent
that care is used in their interpretation. In 2D the vorticity
is restricted to the direction normal to the computational domain,
while in 3D instabilities cause its orientation to wander. 
Thus 2D is useful in situations in which there are physical
reasons to enforce the symmetry (e.g., terrestrial cyclonic storms), but has
nevertheless been used widely in more general applications because of 
computer resource limitations.  The increasing availability of computing clusters 
and software parallelization tools is now making 3D hydrodynamic
simulation more common, 
and we are starting to assess the adequacy and limitations of earlier 2D work.

\par We have calculated a 2D compressible model for 2400 seconds of star time.
We find an average flow velocity of $v_c\approx$ 2.0\mult10$^7$ cm/s
and a convective turnover time 
of $t_c\approx 40$ seconds, so our simulation spans approximately 60 turnover times.
The peak velocity during the course of the simulation is $\sim$5.5\mult10$^7$ cm/s,
corresponding to a peak Mach number of $M\sim0.163$. 
The density fluctuations within the convection zone reach a maximum value of 
$\rho_c'/\lang\rho\rang\sim$ 6\mult10$^{-2}$.
At the convective boundaries the density fluctuations attain a peak value of 
$\rho_b'/\lang\rho\rang\sim$ 12\mult10$^{-2}$.%

\par We find two significant differences between the 2D and 3D models. 
First, we find a significantly decreased turbulent mixing rate in 
the 2D simulation.  Material entrained into the convection zone at the boundaries is 
pulled into the large cyclonic flow patterns in the 2D simulation where large 
composition inhomogeneities persists for several convective turnovers.  In contrast, 
material entrained into the convection zone in the 3D models is homogenized within a 
single convective turnover time.  This effect is illustrated in Figure \ref{composition}, 
which shows the spatial distribution of oxygen abundance, as well as RMS fluctuations 
for both the 2D and 3D simulations.  The 2D simulation retains high level fluctuations
throughout the convective zone, while inhomogeneities in the 3D model are mixed to low 
levels by the time material completes a single crossing.  Comparing Figures \ref{composition}
and \ref{fig_1}, which are snapshots at the same time, show the high entropy oxygen entrained 
at the top boundary corresponds to a negative density perturbation.

\par The second major discrepancy between the 2D and 3D models is the convective
velocity scale.  We find that both the mass averaged convective velocity and the
peak velocity fluctuation are $\sim2$ time larger in the 2D model.
This velocity scale difference may be connected to the lower turbulent mixing 
efficiency in the 2D flow.  We find that the net enthalpy flux, which consists of upward 
and downward directed components, $F_{net} = F_{up} - F_{down}$, is the same 
between the 2D and 3D models. In the 2D model, however, the relative value of the 
individual flux components relative to the net flux, e.g.,  $F_{up}/F_{net}$, 
are much larger than in the 3D model.  Therefore, the 2D model requires a larger velocity
scale to move the same net flux due to the inefficiency of depositing the advected enthalpy
across the convection zone.

\par We find that 2D and 3D models compare well in the wave region, but differ
in the convection zone. The 3D convection is more similar to that of the anelastic
model of \citet{kwg03} and the values predicted by mixing length theory.  
While waves behave similarly in 2D and 3D, turbulent convection does not, particularly
with regards to turbulent mixing efficiency.  Although the spatial resolution of the
2D and 3D models are the same, the number of degrees of freedom in the angular 
direction is much larger in the 3D model, $N_{3D}/N_{2D} = (100)^2/320 \sim $31.
If the number of degrees of freedom were the only important parameter,
one might wonder if 2D would provide a more efficient surrogate to 3D.  It turns out
that 2D is actually more expensive than 3D because the same degrees of freedom in 2D requires
a higher spatial resolution and hence a more severe time step constraint, 
and computational cost is $N_{cost}  \propto N_{space}\times N_{time}$.

\section{CONCLUSIONS}

\par A comparison between the flow properties in fully compressible and anelastic
simulations of stellar oxygen shell burning indicate that the two methods 
produce quantitatively similar results.  Both methods produce convective flows 
in 3D models which are compatible with the results expected for the mixing length
theory of convection for this phase.  The compressible models have been extended to
include additional physics not included in the anelastic model, namely stably stratified
boundary layers and a multi-fluid flow ($N_{species}$ = 25).  The interaction between
the convection and the stable layers excite internal waves which produce larger
thermodynamic fluctuations (up to 11\% in 3D).  Composition inhomogeneities due
to ongoing entrainment events at the convective boundaries also cause density
fluctuations on the several percent level, though material is homogenized rapidly
in the 3D model through turbulent mixing.

\par The relatively large fluctuation which arise at the convective boundaries 
$\sim11$\% may stress the reliability of the anelastic approximation if this region 
is to be included in future simulations of oxygen burning or later epochs, where
entropy and density gradients are large. A variety of convection studies have shown 
that boundary condition type (e.g., hard wall compared to stable layer) alters the 
overall flow pattern within a convection zone \citep{hossain93,rogers05} and therefore
the astrophysically correct conditions should be used.
Low Mach number solvers \citep[e.g.][]{lin06} may be the 
most efficient tools for extending studies of oxygen and silicon shell burning to full 
spherical domains in 3D while retaining the crucial density gradients at the convective
boundaries where convective penetration and entrainment operate, and asymmetric fluctuations 
arise which may have important implications for the evolution of pre-supernova models.
Earlier stages such as carbon and neon burning have 
both milder flows and shallower density gradients and should be better suited for anelastic
methods, even at convective boundaries.  Background expansion and
multi-fluid effects should be included however.  The large time step advantage of the 
anelastic and low Mach number simulations allows for much larger domains or better resolution.

\par Although the efficiency of fully compressible hydrodynamics may be low for the 
Mach numbers modeled, there are no signs that the solver used is breaking down in 
the oxygen shell burning simulations presented here.  This conclusion 
is supported on several grounds, including:
(1) The compressible model is in good agreement quantitatively with the
anelastic methods for the convection zone region, including the velocity scales,
and thermodynamic fluctuation amplitudes, a region in which the anelastic
method is expected to perform well. (2) The compressible simulation of the convection 
zone is also in good agreement with the results of the one-dimensional TYCHO model, 
including the velocity scale and background stratification estimated using mixing length theory.
(3) The dynamics in the stably stratified layers in the simulation agree well
with the analytic solutions to the non-radial wave equation,  including the decomposition of the 
flow into specific, unambiguous modes \citep{ma06a}. 
(4) The fluctuation amplitudes at the convective boundaries which are due to wave motions
are found to be in good agreement with analytic estimates for their scale.

\par Contrary to the assertion made by \citet{almgren06} that compressible
codes should fail for $M < 10^{-2}$, we find a robust solution that agrees with an
anelastic method for the same region simulated.  Additionally, recent compressible
simulations of He shell flash convection by \citet{herwig06} using the finite-volume 
Godunov code RAGE \citep{baltrusaitis96} find a flow with $M\sim10^{-3}$, with 
apparently robust results, including well behaved g-modes.
\citet{almgren06} present an example simulation illustrating the failure of 
PPM to track temperature for a simple flow with Mach number $M\sim0.05$.
This calculation, however, uses a compressible PPM code (FLASH) 
with two major differences from ours (PROMPI): they used the hydrodynamic procedure 
described in \citet{z02} to remove the hydrostatic pressure from the Riemann solver, 
and their stellar model was much more degenerate than ours (the equation of state tends
to become independent of temperature under their conditions, and care
must be taken with cancellation of terms).

\begin{acknowledgements}
  This work was supported in part by 
  the ASCII FLASH center at the University of Chicago. One of us (DA) wishes
  to thank the Aspen Center for Physics for their hospitality.
\end{acknowledgements}

\clearpage

\begin{deluxetable}{lllll}
\tabletypesize{\scriptsize}
\tablewidth{0pt}
\tablecaption{Comparison of Oxygen Burning Models}
\tablehead{  
  \mcol{1}{c}{Variable} &
  \mcol{1}{c}{Units} &
  \mcol{1}{c}{2D-PPM} &
  \mcol{1}{c}{3D-PPM} &
  \mcol{1}{c}{3D-Anelastic\tablenotemark{1}} }
\startdata  
M$_{*}$\tablenotemark{a} & (\msun) & 23 & 23 & 25 \\
M$_{in}$,M$_{out}$\tablenotemark{b} & (\msun) & 1.0(1.5), 2.7(2.4) & 1.0(1.5), 2.7(2.4) & 1.2, 2.3 \\
r$_{in}$,r$_{out}$\tablenotemark{b} & (10$^9$ cm) & 0.3(0.44), 1.0(0.85) & 0.3(0.44), 1.0(0.85) & 0.45, 0.84 \\
L$_{net}$\tablenotemark{c} & (erg/s) & 3.2\mult 10$^{46}$ & 3.5\mult 10$^{46}$ & 1.5\mult 10$^{45}$\\
$\Delta\theta$,$\Delta\phi$ & (deg.) & 90 & 30, 30 & 360, 180 \\
Zones/Modes  & ($n_r$\mult $n_{\phi}$\mult $n_{\theta}$) &  
400\mult320\mult1 &  400\mult100\mult100 & 145\mult63($l$)\mult31($m$) \\
$v_{\hbox{max}}$& (10$^7$ cm/s) &  7.2 & 3.8 & 1.8\\
$\langle v_{\hbox{\tiny conv}}\rangle$ &(10$^7$ cm/s) & 2.0 & 0.8 & 0.49\\
$M_{peak}$& - & 0.163 & 0.09 & $\sim$0.04\\
$M_{rms}$ & - & 0.03 & 0.01 & $\sim$0.01 \\
t$_{\hbox{\tiny conv}}$ &(s)  & 40 & 103 & 159 \\
t$_{\hbox{\tiny max}}$ &(s) & 2400 & 800  & 6500\\
max\{ $\rho'_c/\langle\rho\rangle$ \}\tablenotemark{d} & 10$^{-2}$& 1.0(6.0)\tablenotemark{e} & 0.5 & 0.2\\
max\{ $T'_c/\langle T\rangle$ \}  &10$^{-2}$&  0.25 & 0.05 & 0.06\\
max\{ $\rho'_{b}/\langle\rho\rangle$ \}  & 10$^{-2}$ & 12.0 & 11.0 & - \\
max\{ $T'_{b}/\langle T\rangle$ \}& 10$^{-2}$& 2.7 & 1.0 & - \\

\enddata

\tablenotetext{1}{The values quoted for the 3D anelastic model are from \cite{kwg03}.}
\tablenotetext{a}{Zero age main sequence mass.}
\tablenotetext{b}{Values in parentheses indicate the extents of the convection
  zone for the compressible models and the other values indicate the extents of the entire
  computation domain including the stable layers.}
\tablenotetext{c}{The net luminosity for the compressible models,
  L$_{net} = \int (\epsilon_{nuc}+\epsilon_{\nu\bar{\nu}}) dM$,  is estimated at t$\sim$400 s
  and is slowly decreasing with time due to an overall background expansion occuring within the burning region.}
\tablenotetext{d}{The  {\em c} and {\em b} subscripts indicate thermodynamic fluctuation amplitudes 
  that are estimated in the convection zone and in the region of the convective boundary, respectively.}

\tablenotetext{e}{The value in parentheses is the maximum density fluctuation estimated over two convective
  turnovers while the other value represents the time averaged maximum fluctuation. The significantly larger
  fluctuations seen in the 2D model compared with those in the 3D model occur near the center of large
  vortices which persist for several convective turnover times in the 2D model but are absent in the 3D
  convective flow.
}
\label{tab_1}
\end{deluxetable}

\clearpage

\end{document}